\documentstyle[twocolumn,aps,prl,epsf]{revtex}
\begin{document}
%****
%\preprint{NUSc/00-05}
\title{Chaos in a Jahn-Teller Molecule}

\author{R.S. Markiewicz}

\address{Physics Department and Barnett Institute, 
Northeastern U.,
Boston MA 02115}
\maketitle

\begin{abstract}
The Jahn-Teller system $E\otimes b_1\oplus b_2$ has a particular degeneracy,
where the vibronic potential has an elliptical minimum.  In the general case
where the ellipse does not reduce to a circle, the classical motion in the
potential is chaotic, tending to trapping near one of the extrema of the 
ellipse.  In the quantum problem, the motion consists of correlated tunneling
from one extremum to the opposite, leading to an average angular momentum
reminiscent of that of the better known $E\otimes e$ dynamic Jahn-Teller system.
\end{abstract}


%****
\narrowtext
%****

In the well-known $E\otimes e$ Jahn-Teller (JT) effect, a molecule has a 
two-fold electronic degeneracy coupled to a doubly degenerate vibrational mode.
This leads to a `conical intersection' in the vibronic potential which has a 
degenerate, circular minimum (`Mexican hat potential'), although higher-order 
vibronic coupling can break the ring up into three degenerate minima along the 
trough (`tricorn potential')\cite{JT1}.  Quantum mechanically, the coupled 
electron-molecular vibrational (vibronic) wave function can tunnel between the 
three minima, leading to a ground state with a net angular momentum\cite{HLH}.  
Remarkably, this `orbital' angular momentum is quantized in half-integer 
multiples of $\hbar$, indicating the strong coupling of electronic and molecular
motions.  This quantization is a signature of the Berry phase\cite{CAM,Ber} of 
$\pi$ associated with the dynamic Jahn-Teller effect; the $\pi$ Berry phase has 
been experimentally verified in triangular Na$_3$ molecules\cite{Del}. Points of
conical intersection lead to {\it chaotic} behavior in the vibrational spectra,
manifested quantum mechanically by anomalous level statistics\cite{chao}.
However, the high symmetry of the $E\otimes e$ problem precludes 
chaos\cite{noch}, so multimode interactions must be included, and the chaos
generally appears at high energies (above the conical intersection) where many
vibrational modes are excited.

\begin{figure}
\leavevmode
   \epsfxsize=0.3\textwidth\epsfbox{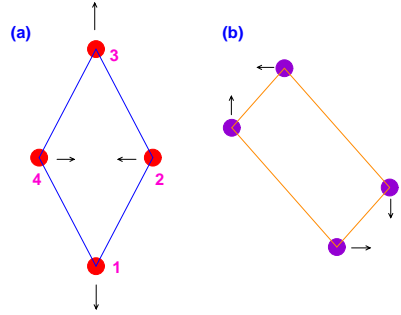}
\vskip0.5cm 
\caption{$B_1$ (a) and $B_2$ (b) distortions of a square X$_4$ molecule.}
\label{fig:1}
\end{figure}

Here, it is shown that a simple modification of the symmetry preserves the 
anomalous Berry phase, yet leads to chaotic behavior at much lower energies,
without the need of additional mode coupling.  This case is the square
X$_4$ molecule with square planar symmetry, $D_{4h}$, corresponding 
to an $E\otimes b_1\oplus b_2$ Jahn-Teller problem\cite{JT1}, Fig.~\ref{fig:1}.
The high symmetry allows two JT modes, with independent frequencies $\omega_i$,
$i=1,2$, and electron-vibration couplings $V_i$.  In the special case $\omega_1
=\omega_2$, $V_1=V_2$, the problem reduces exactly to that of the $E\otimes e$ 
molecule.  However, there is an intermediate case, which seems not to have been
explored till now.  When $V_1/\omega_1=V_2/\omega_2$, the two modes have the
same JT stabilization energy, $E_{JT}^{(i)}=V_i^2/2M\omega_i^2$, and hence the
vibronic potential has an elliptic minimum, which is not circular unless
$\omega_1=\omega_2$.  Given the elliptic minimum, the possibility of a periodic
orbit arises.  However, angular momentum is not conserved.  In the present
paper I analyze the resulting motion.

The phonon modes $B_i$ of amplitude $Q_i$ are defined as follows.  
The four atomic positions, Fig.~\ref{fig:1}a, can be written as 
\begin{eqnarray}
\vec r_1=\vec r_{10}-Q_1\hat y+Q_2\hat x,
\nonumber \\
\vec r_2=\vec r_{20}-Q_1\hat x-Q_2\hat y, 
\nonumber \\
\vec r_3=\vec r_{30}+Q_1\hat y-Q_2\hat x, 
\nonumber \\
\vec r_4=\vec r_{40}+Q_1\hat x+Q_2\hat y,
\label{eq:5a}
\end{eqnarray}
where the $\vec r_{i0}$'s are the positions of the X atoms in the undistorted 
square.  The vibronic interaction Hamiltonian is
\begin{eqnarray}
H_{vib}=V_1Q_1T_x+V_2Q_2T_y
\nonumber \\
=V_1Q_1\left(\matrix{1&0\cr 0&-1}\right)
+V_2Q_2\left(\matrix{0&1\cr 1&0}\right).
\label{eq:5}
\end{eqnarray}
Here the electronic operators are represented by the pseudospin $T_i$'s and 
other factors are included in the electron-phonon coupling $V_i$.  
To the vibronic Hamiltonian must be added an electronic term $H_{el}$ and a 
phononic part $H_{ph}$, with 
\begin{eqnarray}
H_{ph}={1\over 2M}\bigl(P_1^2+P_2^2+M^2\omega_1^2Q_1^2+M^2\omega_2^2Q_2^2\bigr),
%\nonumber \\
\label{eq:5b}
\end{eqnarray}
with $\omega_i$ the bare phonon frequencies.
A spin-orbit coupling can be included\cite{Bal} 
\begin{equation}
H_{so}=\lambda\vec L\cdot\vec S.
\label{eq:6b}
\end{equation}

For a static JT effect, the momenta $P_i$ can be neglected, and the $Q_i$ are
chosen to minimize the energy, Eqs.~\ref{eq:5},\ref{eq:5b}.  The solution can be
written in terms of the JT energy $E_{JT}^{(i)}=V_i^2/(2M\omega_i^2)$.  For
$E_{JT}^{(1)}\ne E_{JT}^{(2)}$, the lowest energy state consists of a distortion
of the mode with larger JT energy only.  For instance, if $E_{JT}^{(2)}>E_{JT}^
{(1})$, the solution is $Q_1=0$, $Q_2=V_2/(M\omega_2^2)$, $E=-E_{JT}^{(2)}$.

\begin{figure}
\leavevmode
   \epsfxsize=0.3\textwidth\epsfbox{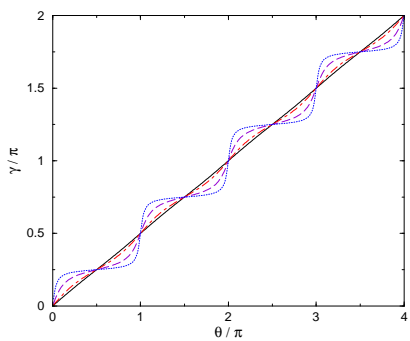}
\vskip0.5cm 
\caption{Electronic phase $\gamma$ vs phononic phase $\theta$, for $\delta$ = 
0.1 (solid line), 1 (dotdashed line), 10 (dashed line), and 100 (dotted line).}
\label{fig:4}
\end{figure}

Special cases arise when 
\begin{equation}
E_{JT}^{(1)}=E_{JT}^{(2)}\equiv E_{JT}. 
\label{eq:2d}
\end{equation}
Eliminating the electrons produces the vibronic potential surfaces 
\begin{equation}
E_{\pm}={M\over 2}\left(\omega_1^2Q_1^2+\omega_2^2Q_2^2\right)\pm\sqrt{
V_1^2Q_1^2+V_2^2Q_2^2+{\lambda^2\over 4}}.  
\label{eq:6c}
\end{equation}
When Eq.~\ref{eq:2d} is satisfied, the lower vibronic surface has a minimum 
which is degenerate along a trough, similar to the Mexican hat:
\begin{eqnarray}
Q_1^0=Q_0^0{\cos{\theta}\over\omega_1},
\nonumber\\
Q_2^0=Q_0^0{\sin{\theta}\over\omega_2},
\label{eq:6d}
\end{eqnarray}
with $Q_0^0=\sqrt{2E_{JT}-\lambda^2/8E_{JT}}$, and $\theta$ arbitrary.
Near the trough, the lower potential surface can be expanded:
\begin{equation}
E_{-}={M\over 2}\alpha\left(\vec g\cdot\vec q\right)^2,
\label{eq:6e}
\end{equation}
with $\vec q=(q_1,q_2),$ $q_i=Q_i-Q_i^0$, $\alpha =1-\lambda^2/16E_{JT}^2 $
and $\vec g=(\omega_1\cos{\theta}, \omega_2\sin{\theta})$ -- that is, there is 
a restoring force only `perpendicular' to the trough.
Defining $\beta_{\omega}=\omega_2/\omega_1$, the electronic eigenvectors are
\begin{eqnarray}
\psi_+=\cos\gamma\psi_1+\sin\gamma\psi_2 \nonumber\\
\psi_-=-\sin\gamma\psi_1+\cos\gamma\psi_2, 
\label{eq:10}
\end{eqnarray}
where $\tan\gamma =(\sqrt{1+\delta\sin^2\theta}-\cos\theta )/
(\beta_{\omega}\sin\theta)$
and $\delta =\beta_{\omega}^2-1$, Fig.\ref{fig:4}.  By convention, $\omega_2$
is assumed to be the higher frequency ($\beta_{\omega}\ge 1$).

If additionally $\omega_1=\omega_2$, the problem reduces to the $E\otimes e$ 
problem, and in Eq.~\ref{eq:10} $\gamma =\theta /2$: {\it the electronic wave 
function is double valued}: when $\theta$ changes by $2\pi$, $\gamma$ has only 
changed by $\pi$ (the wave functions have changed sign).  This sign change is 
the signature of a Berry phase\cite{CAM,Ber}, and causes the vibronic orbital 
angular momentum to take on half-integer values\cite{HLH}.  
This can be seen as follows\cite{JT1}.  The z-component of orbital angular 
momentum is $L_z=(Q_1P_2-Q_2P_1)/\hbar$, and the operator which commutes with 
the vibronic hamiltonian, Eq.~\ref{eq:5} is
\begin{equation}
j_z=L_z+{1\over 2}T_z.
\label{eq:9a}
\end{equation}
Since $L_z$ is quantized in integers, $j_2$ has half-integral quanta.  Note from
Fig.~\ref{fig:4} that even when $\omega_1\ne\omega_2$, $\theta$ must change by
$4\pi$ to produce a $2\pi$ change in $\gamma$, suggesting a similar Berry phase.
%%%%NEW
This can be directly demonstrated.  The Berry phase is\cite{ZG}
\begin{eqnarray}
\gamma_B=-s\int_0^{2\pi}{\partial\gamma\over\partial\theta}d\theta
\nonumber \\
=-\beta s\int_0^{2\pi}{d\theta\over 1+\sin^2{\theta}}=-2\pi s,
\label{eq:9b}
\end{eqnarray}
where $s$ is half an odd integer, introduced to make the total wave function
single valued.  Thus the Berry phase is $\pi$, modulo $2\pi$, for any 
anisotropy.
%%%%NEW

While this is a standard JT problem, I have not found any detailed analysis of
the limit Eq.~\ref{eq:2d}.  As a first step, I perform a canonical 
transformation
\begin{equation}
H'=e^{iS}He^{-iS}=H+i[S,H]-...
\label{eq:30}
\end{equation}
with
\begin{equation}
S=-\bigl({V_1\over\omega_1^2}P_1T_x+{V_2\over\omega_2^2}P_2T_y\bigr).
\label{eq:31}
\end{equation}
The canonical transformation can be performed exactly\cite{Wag}, but for present
purposes only the first order result is needed.  S, Eq.~\ref{eq:31}, was chosen 
to exactly cancel the term linear in $Q$.  It yields a correction
\begin{equation}
H_2'=i[S,H_{vib}]=-{T_0\over 2}\bigr({V_1^2\over\omega_1^2}+{V_2^2\over\omega_2
^2}\bigr)+{V_1V_2\over\omega_1^2\omega_2^2}AT_z,
\label{eq:32}
\end{equation}
with $T_0$ the identity matrix,
\begin{equation}
A=\omega_2^2P_1Q_2-\omega_1^2P_2Q_1=-\omega_+^2L_z+\omega_-^2(P_1Q_2+P_2Q_1),
\label{eq:33}
\end{equation}
and $\omega_{\pm}^2=(\omega_1^2\pm\omega_2^2)/2$. Thus,
when $\omega_-=0$, $H_2'$ is proportional to $L_zT_z$, and the angular momentum
$j_z=L_z+T_z/2$ is conserved (Eq.~\ref{eq:9a}).  For the present case $\omega_-
\ne 0$ and $j_z$ is not constant.  

Given the presence of a circular trough in 
the potential, circulating orbits should be possible: could it be that there is 
a nonvanishing average $<j_z>\ne 0$ even though $j_z$ is not constant?  This 
possibility can be explored in the related classical Hamiltonian (particle in a 
non-linear potential well) by numerically integrating the equations of motion
\begin{eqnarray}
\ddot Q_i=-{dE_-\over dQ_i}=-\omega_i^2Q_i\bigl(1-{2E_{JT}\over\sqrt{V_1^2Q_1^2
+V_2^2Q_2^2+\lambda^2/4}}\bigr) 
\nonumber\\
\simeq -\alpha\omega_i^2Q_i\bigl(1-{q_0\over\sqrt{\omega
_1^2Q_1^2+\omega_2^2Q_2^2}}\bigr) 
\label{eq:34}
\end{eqnarray}
where the last form utilizes the quadratic approximation, Eq.~\ref{eq:6e}, 
dots indicate time derivatives, and $q_0^2=2E_{JT}\alpha$.  The integral is
evaluated using a Runge-Kutta routine with initial conditions $\vec Q(0)=
(q_0/\omega_1,0)$, $\dot{\vec Q}(0)=(0,\beta q_0/\omega_2)$.  In the remaining 
analysis, I take $\lambda =0$.  

\begin{figure}
\leavevmode
   \epsfxsize=0.4\textwidth\epsfbox{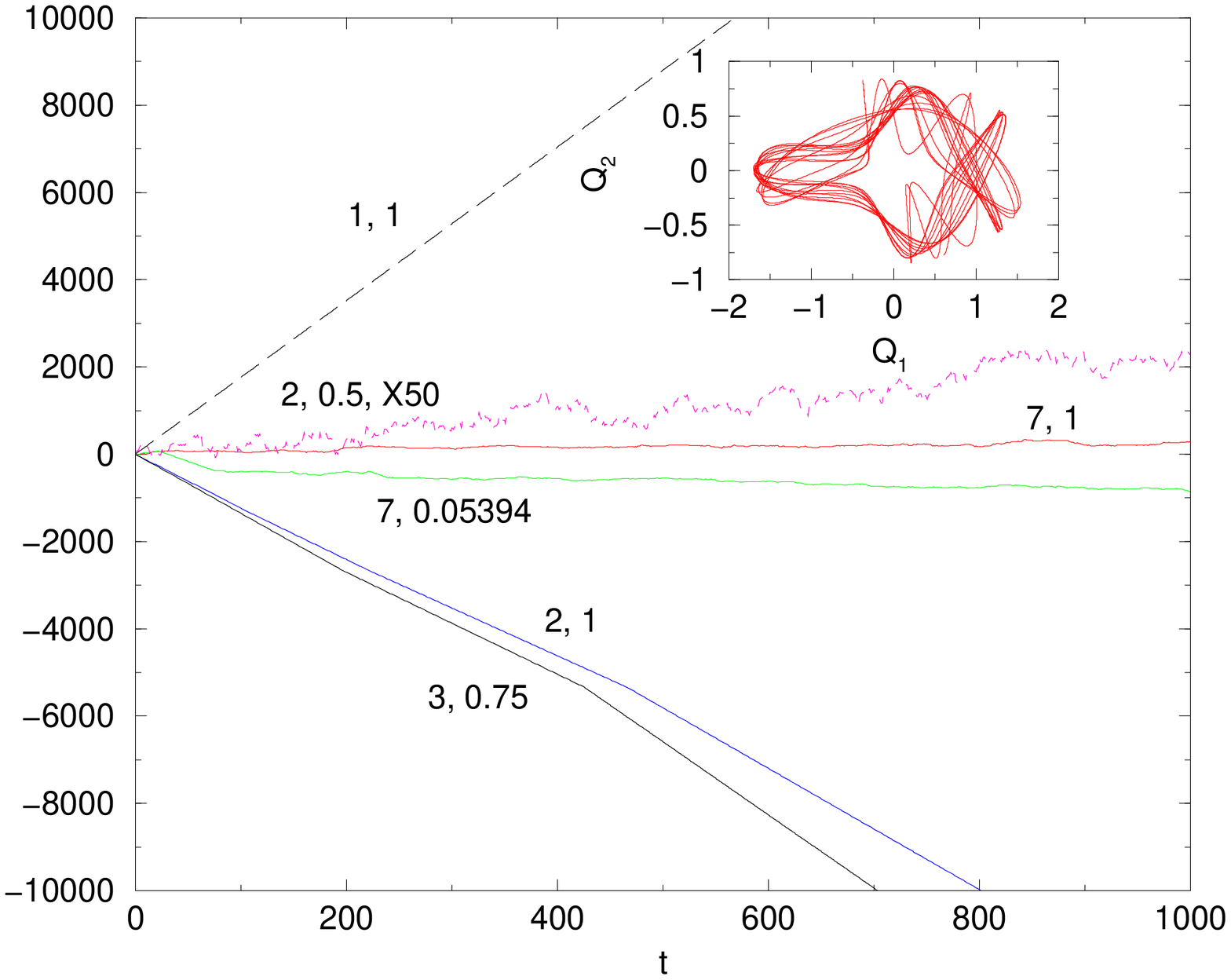}
\vskip0.5cm 
\caption{Winding angle $\phi$ calculated from Eq.~\protect\ref{eq:35}, for
several values of $(\beta_{\omega},\beta )$.
%$\omega_2/\omega_1$: solid line, $\omega_2/\omega_1 =7$; 
%dashed line = $\phi /100$ for $\omega_2/\omega_1=1$.  
Inset = time series, $Q_1,Q_2(t)$ for $\beta_{\omega}=\omega_2/\omega_1 =2$,
$\beta =0.5$.}
\label{fig:41}
\end{figure}

Given $Q_1(t)$, $Q_2(t)$, a winding angle $\phi$ is defined such that
\begin{equation}
\dot\phi ={Q_1\dot Q_2-Q_2\dot Q_1\over Q_1^2+Q_2^2}.
\label{eq:35}
\end{equation}
If one applies this procedure to the $E\otimes e$ problem ($\omega_2=\omega_1$),
the results are quite simple (long dashed line in Fig.~\ref{fig:41}): 
%Eq.~\ref{eq:34} has a solution $|Q|=q_0/\omega_1$, $\phi =\beta t$.  
$\phi$ increases linearly with time, although the frequency is not constant, but
varies approximately logarithmically with the velocity parameter $\beta$.  By
contrast, when $\omega_2\ne\omega_1$ Figure~\ref{fig:41} shows that $\phi$ is 
generically a random function of time, with no linearly increasing part 
indicative of a non-zero $<j_z>$.  The various data sets are characterized by
the two parameters $(\omega_2 /\omega_1,\beta )$. (The figure utilizes the exact
form of Eq.~\ref{eq:34}; the approximate form yields equivalent results.)   
The figure also clearly suggests that the motion is chaotic.  This is further
indicated by the direct time series, inset of Fig.~\ref{fig:41}. 

\begin{figure}
\leavevmode
   \epsfxsize=0.44\textwidth\epsfbox{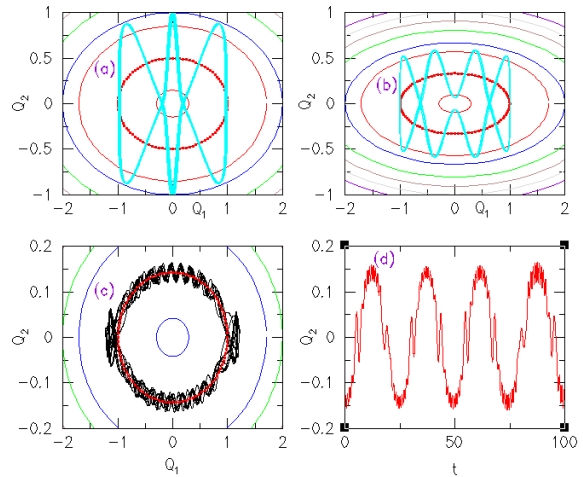}
\vskip0.5cm 
\caption{Time series, $Q_2(t)$ vs $Q_1(t)$ (or vs t, in (d)) for several
choices of $\beta_{\omega},\beta )$: (a) = (2,1), (b) = (3,0.75), (c,d) = 
(7,0.05394).  In frames (a-c), the ellipses are equipotential contours, with the
beaded contour representing the potential minimum.}
\label{fig:43}
\end{figure}

\begin{figure}
\leavevmode
   \epsfxsize=0.44\textwidth\epsfbox{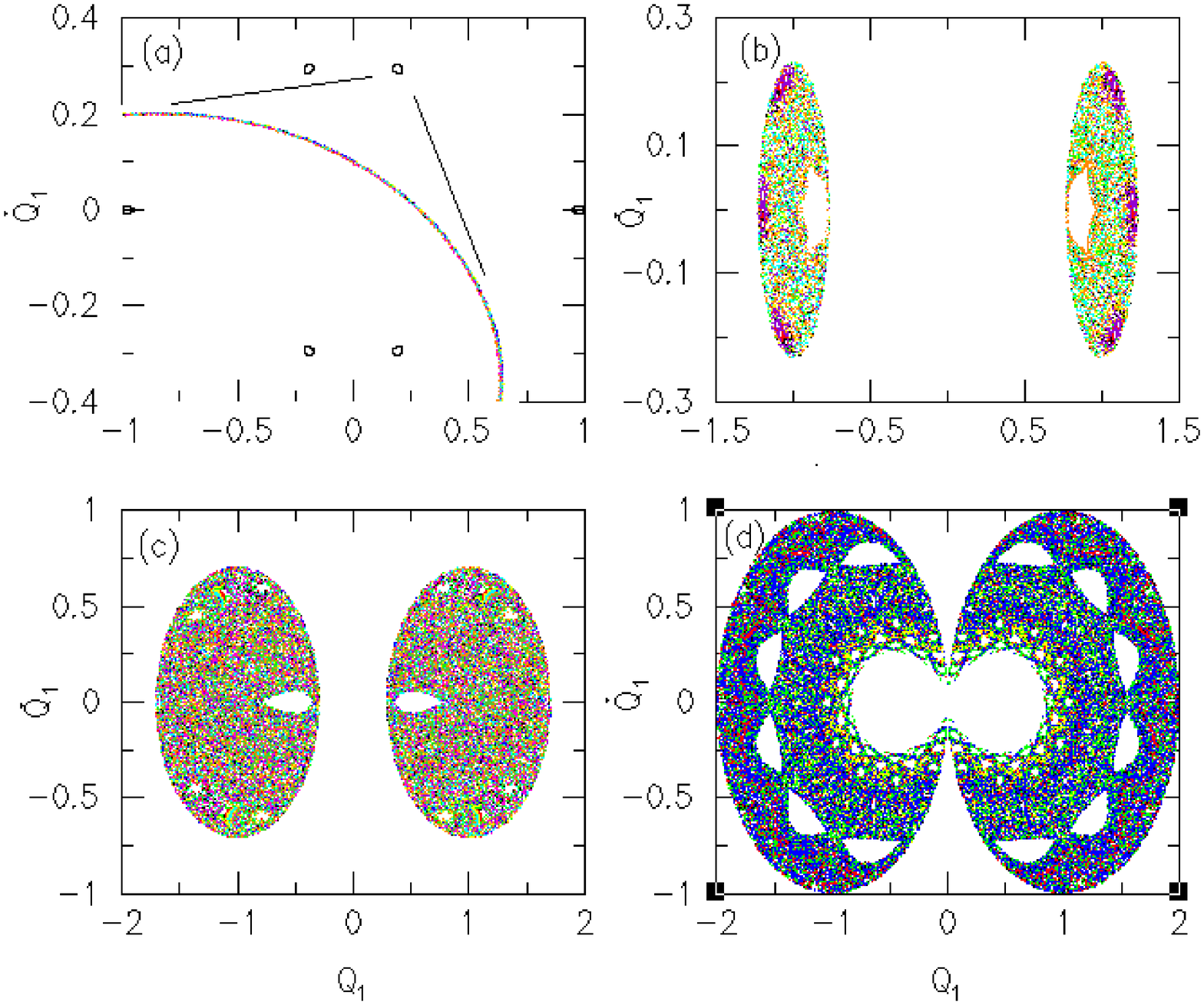}
\vskip0.5cm 
\caption{Poincare maps for $(\beta_{\omega},\beta )$ = (2,1) (a), (7,0.05394)
(b), (2,0.5) (c), (7,1) (d).  In (a), one attractor is shown blown up.}
\label{fig:44}
\end{figure}

On the other hand, there are certain special values of the initial conditions
for which the motion is approximately periodic, and $\phi$ does increase 
linearly with time.  These values may most easily be found by plotting $\phi
(T)$ vs $\beta$ for some long time $T$.  Typical examples are illustrated in
Fig.~\ref{fig:41}, while the time series are shown in Fig.~\ref{fig:43}.  
Poincare maps (plots of $Q_1$ vs $\dot Q_1$ when $Q_2=0$), 
Figure~\ref{fig:44}, confirm the chaotic nature.  (Note that the curve $(7,
0.05394)$ is almost periodic -- see particularly $Q_1(t)$, Fig.~\ref{fig:43}d 
-- but the Poincare map is clearly chaotic, Fig.~\ref{fig:44}d.)  While the 
$E\otimes e$ limit, $\beta_{\omega}=1$, is quasiperiodic (the Poincare map is a 
smooth closed curve), for $\beta_{\omega}\ne 1$ even the special values are
weakly chaotic, with the Poincare maps, Fig.~\ref{fig:44}a, having a finite
spread away from smooth curves.  The similarity of these special trajectories to
scars in e.g., Sinai stadia\cite{Sin} should be noted.  

How is this chaotic behavior manifested in the quantum limit?  To explore this,
it is convenient to first rescale the variables, so that the potential has 
circular symmetry, and the anisotropy appears in the ionic mass, $m_i=M(\omega_0
/\omega_i)^2$, with $\omega_0^2=(\omega_1^2+\omega_2^2)/2$, and then reduce the 
problem to one dimension by assuming that the motion is 
confined to the bottom of the trough and only $\phi$ varies.  The
Hamiltonian becomes $H=-\hbar^2h/(2m_+\rho_0^2)$, where $\rho_0$ is the 
equillibrium trough radius, $m_{\pm}^{-1}=(m_2^{-1}\pm m_1^{-1})/2$ and
\begin{equation}
h=\partial_{\phi}^2+\hat\alpha [\cos{2\phi}({3\over 2}-\partial_{\phi}^2)+3\sin{
2\phi}\partial_{\phi}]-A_4\cos{4\phi},
\label{eq:36}
\end{equation}
with $\hat\alpha =m_+/m_-=(\beta_{\omega}^2-1)/(\beta_{\omega}^2+1)$
and higher order vibronic effects are incorporated in the term proportional to
$A_4$.
Schroedinger's equation can be integrated numerically, letting $\psi (\phi ,t)=
\psi (j\epsilon ,n\delta )\equiv\psi_j^n$, with $\partial_{\phi}\psi =(\psi_{j+1
}^n-\psi_j^n)/\epsilon$, and\cite{QCh}
\begin{equation}
\psi_j^{n+1}=e^{-iH\delta /\hbar}\psi_j^n\simeq{1-iH\delta /2\hbar\over 1+iH
\delta /2\hbar}\psi_j^n,
\label{eq:37}
\end{equation}
or finally $(1-i\gamma h)\psi_j^{n+1}=(1+i\gamma h)\psi_j^n$, with $\gamma =
\hbar\delta /4m_+\rho_0^2$.
\begin{figure}
\leavevmode
   \epsfxsize=0.33\textwidth\epsfbox{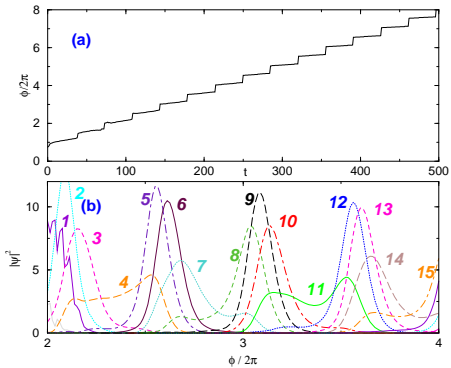}
\vskip0.5cm 
\caption{Quantum time evolution, showing (a) position of wave function peak as a
function of time, and (b) actual distribution of $|\psi |^2$ at several 
equally-spaced time intervals ($\hat\alpha=0.3$, $A_4=0$).}
\label{fig:45}
\end{figure}

\begin{figure}
\leavevmode
   \epsfxsize=0.42\textwidth\epsfbox{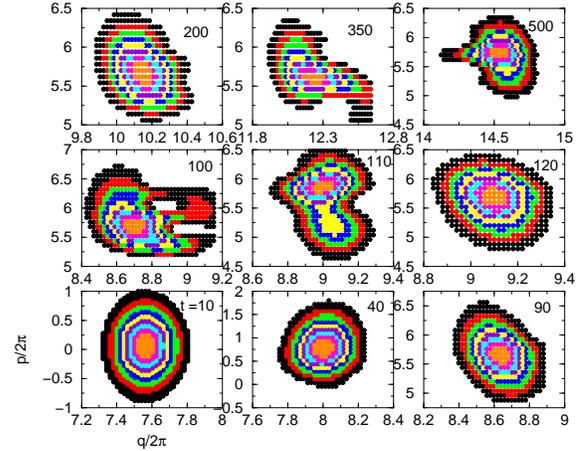}
\vskip0.5cm 
\caption{Contour plot of Husimi distribution $\rho_H$ of data similar to that of
Fig.~\protect\ref{fig:45} at several time intervals.  An interwell hopping event
occurs between times 90 and 120.}
\label{fig:46}
\end{figure}

Equation \ref{eq:37} was integrated numerically, assuming an initial gaussian
distribution.  Figure \ref{fig:45}b shows $|\psi (\phi ,t)|^2$ for a variety of 
times $t$.  The data can be better understood from Fig. \ref{fig:45}a, which
plots $\phi_{max}$ vs $t$, where $\phi_{max}$ is that value of $\phi$ for which
$|\psi |^2$ has its maximum value.  The wave function remains trapped in one of
the effective potential wells, then quickly hops to the next one in a relatively
short time.  This hopping takes place by the probability spreading over two
adjacent wells, as shown in Fig.~\ref{fig:45}b at times 4, 7, and 11. 
The tunneling is coherent, so there is a net circulation.
Additional information can be found by analyzing the Husimi density\cite{Hus}
$\rho_H(p,q)=|<p,q|\psi >|^2$, with
\begin{equation}
<p,q|\psi >=\root 4 \of {{s\over\hbar\pi}}\int exp[-{s(\phi -q)^2\over 2\hbar}
-i{p\over\hbar}(\phi -{q\over 2})]\psi (\phi )d\phi ,
\label{eq:38}
\end{equation}
which describes the approximate smearing of $\psi$ in $q$ and $p$ as a function
of time.  Typical results are shown in Fig. \ref{fig:46}, for squeezing 
parameter $s=1$.

Thus, the quantum system shows a `memory' of the classical chaos, in that the
wave function shows similar trapping near the points $Q_2=0$.  However, 
whereas the wavefunctions appear to vary stochastically from cycle to cycle,
Fig.~\ref{fig:45}b, the average of the wave function progresses smoothly, 
Fig.~\ref{fig:45}a.  The main difference is that classically, the wave function
can be reflected from a trapping region, reversing its direction of motion, 
while the quantum wave function always moves in the same direction, similar to 
the classical problem with special initial conditions.  It seems plausible to 
interpret the special choice of initial conditions as analogous to a 
Bohr-Sommerfeld quantization condition in the quantum problem.  

As shown in Fig.~\ref{fig:45}a, the position of the wave function peak has a
step-like component superposed on an average shift with time.  This average
shift is independent of the mass anisotropy\cite{MKI}, hence corresponding to
the {\it same quantized angular momentum} as in the isotropic case.  This is
consistent with the Berry phase remaining $\pi$, Eq.~\ref{eq:9b}, but somewhat
surprising in light of the classical chaos.  Blumel\cite{Blu} has suggested that
this might be a manifestation of quantum {\it localization in angular momentum 
space}\cite{Qloc}, 
while the classical problem leads to angular momentum space diffusion.  This
possibility will be explored in future work.

In addition to its molecular interest, the present results may have condensed 
matter applications.  Berryonic matter\cite{Berr} has been postulated to
explain anomalous properties of Buckyballs and other dynamic JT systems, but
based on unit cells of triatomic molecules.  Potential applications are greatly
expanded for bases of square molecules\cite{MKI}.

I thank J. Jose, F.S. Ham, R. Englman, R. Blumel, and B. Barbiellini for 
stimulating conversations.


\begin{references}
\bibitem{JT1}I.B. Bersuker and V.Z. Polinger, ``Vibronic Interactions in 
Molecules and Crystals" (Springer, Berlin, 1989); M.D. Kaplan and B.G. Vekhter, 
``Cooperative Phenomena in Jahn-Teller Crystals" (Plenum, N.Y., 1995).
\bibitem{HLH}G. Herzberg and H.C. Longuet-Higgins, Disc. Farad. Soc. {\bf 35},
77 (1963).
\bibitem{CAM}C.A. Mead, Rev. Mod. Phys. {\bf 64}, 51 (1992).
\bibitem{Ber}M.V. Berry, Proc. Roy. Soc. London, A{\bf 392}, 45 (1984).
\bibitem{Del}H. von Busch, Vas Dev, H.-A. Eckel, S. Kasahara, J. Wang, W.
Demtr\"oder, P. Sebald, and W. Meyer, Phys. Rev. Lett. {\bf 81}, 4584 (1998).
%G. Delacr\'etaz, E.R. Grant, R.L. Whetten, L. W\"oste, and J.W.
%Zwanziger, Phys. Rev. Lett. {\bf 56}, 2598 (1986).
\bibitem{chao}E. Haller, H. K\"oppel, and L.S. Cederbaum, J. Molec. Spectrosc.
{\bf 111}, 377 (1985); A. Delon, R. Jost, and M. Lombardi, J. Chem. Phys. {\bf
95}, 5701 (1991).
\bibitem{noch}H. K\"oppel, W. Domcke, and L.S. Cederbaum, Advanc. Chem. Phys. 
{\bf 57}, 59 (1984).
\bibitem{Bal}C.J. Ballhausen, Theoret. Chim. Acta (Berl.) {\bf 3}, 368 (1965). 
\bibitem{ZG}J.W. Zwanziger and E.R. Grant, J. Chem. Phys. {\bf 87}, 2954 (1987).
\bibitem{Wag}M. Wagner, in ``The Dynamical Jahn-Teller Effect in Localized
Systems'', ed. by Yu.E. Perlin and M. Wagner (North Holland, Amsterdam, 1984),
p. 155.
\bibitem{Sin}S. Sridhar, Phys. Rev. Lett. {\bf 67}, 785 (1991); A. Kudrolli, S. 
Sridhar, A. Pandey, and R. Ramaswamy, Phys. Rev. E{\bf 49}, 11 (1994).
\bibitem{QCh}H.J. Korsch and H. Wiescher, in ``Computational Physics'', edited
by K.H. Hoffman and M. Schreiber (Springer, Berlin, 1996), p. 225.
\bibitem{Hus}K. Husimi, Proc. Phys. Math. Soc. Jpn. {\bf 22}, 264 (1940). 
\bibitem{Berr}N. Manini, E. Tosatti, and S. Doniach, Phys. Rev. B{\bf 51}, 
3731 (1995).
\bibitem{MKI}R.S. Markiewicz and C. Kusko, unpublished.
\bibitem{Blu}R. Blumel, personal communication.
\bibitem{Qloc}F. Borgonovi, G. Casati, and B. Li, Phys. Rev. Lett. {\bf 77},
4744 (1996); K.M. Frahm and D.L. Shepelyansky, Phys. Rev. Lett. {\bf 78}, 1440
(1997).
\end{references}
\end{document}